# Large-scale five- and seven-junction epitaxial graphene devices

Dinesh Patel,[1,2] Martina Marzano,[1,3,4] Chieh-I Liu,[1] Heather M. Hill,[1] Mattias Kruskopf,[1,5] Hanbyul Jin,[1,5] Jiuning Hu,[1,5] David B. Newell,[1] Chi-Te Liang,[2] Randolph Elmquist,[1] and Albert F. Rigosi[1,a)]

[1]Physical Measurement Laboratory, National Institute of Standards and Technology (NIST), Gaithersburg, Maryland, 20899-8171, USA

[2]Department of Physics, National Taiwan University, Taipei, 10617, Taiwan

[3]Department of Electronics and Telecommunications, Politecnico di Torino, Torino, 10129, Italy

[4]Istituto Nazionale di Ricerca Metrologica, Torino, 10135, Italy

[5]Joint Quantum Institute, University of Maryland, College Park, MD 20742, USA

The utilization of multiple current terminals on millimeter-scale graphene *p-n* junction devices has enabled the measurement of many atypical, fractional multiples of the quantized Hall resistance at the $\nu = 2$ plateau ($R_\mathrm{H} \approx 12906$ $\Omega$). These fractions take the form $\frac{a}{b} R_\mathrm{H}$ and can be determined both analytically and by simulations. These experiments validate the use of either the LTspice circuit simulator or the analytical framework recently presented in similar work. Furthermore, the production of several devices with large-scale junctions substantiates the approach of using simple ultraviolet lithography to obtain junctions of sufficient sharpness.

---

Graphene, composed of carbon atoms arranged in two-dimensional honeycomb lattice, has been extensively studied more than a decade, in part because of its excellent optical, mechanical and electrical transport properties [1-4]. The quantum Hall effect (QHE) in graphene gives resistance values at $\frac{1}{(4n+2)}\frac{h}{e^2}$, where $n$ is an integer, $h$ is the Planck constant, and $e$ is the elementary charge. Graphene $p$-$n$ junctions ($pn$Js), which are suitable for one to explore transport in the QHE [5-18], enable one to access various multiples and fractions of the von Klitzing constant. These types of graphene devices also have additional applications in electron optics [19-22], 2D materials [23-27], and quantum Hall resistance standards [28-38].

For clarity, a $pn$J device contains some form of interface at which a positively-doped and negatively-doped region meet. For graphene, whose Fermi level can be electrically or chemically modulated, such an interface can be effectively one-dimensional, allowing edge state electrons to tunnel from one region to the other. This behavior results in the observation of quantized longitudinal resistances due to the presence of the junction. Typically, these devices are of sub-millimeter sizes due to constraints on top-gating. One motivation for pursuing large-scale $pn$J devices is to determine the feasibility of using quantum transport across the junctions to access different quantized values of resistance, as shown in previous studies [39-41]. One first major hurdle is to fabricate large-scale devices without the need for top-gating, since such techniques typically increase in difficulty as the device incorporates more elements. Though extensive analyses exist on Landauer-Büttiker edge state equilibration [5-8, 42-46], creating a $pn$J device capable of accessing different plateaus with top gates is a difficult task. Instead, one approach to accessing different quantized values is to incorporate multiple current terminals, which opens the parameter space within which $pn$J devices are able to be operated.

For millimeter-scale device fabrication, epitaxial graphene (EG) is grown to accommodate device size, but the issue of processing the correspondingly large $pn$Js was not trivial, as shown in previous work [47]. This work elaborates on further efforts involving the use of standard ultraviolet photolithography (UVP) and ZEP520A to build $pn$Js having widths smaller than 200 nm. Devices were verified via quantum Hall transport measurements and LTspice current simulations [48], and multiple current terminals and configurations were used to test the viability of the simulations as well as the quality of the devices. Furthermore, analytical methods recently reported were also used to predict atypical fractions of the quantized Hall resistance, $R_H$, that would become experimentally accessible depending on the configuration of the current terminals [49]. These experiments also serve as supporting evidence on the validity of those analytical methods, which provide easily implementable algorithms for determining effective quantized resistances in complicated $pn$J circuits.

Simulations for the *pn*J devices were performed with the analog electronic circuit simulator LTspice in an identical manner as demonstrated for similar devices in other works [47, 49-51]. The circuit uses both *p*-type and *n*-type *k*-terminal quantum Hall elements, designated as either having ideal counterclockwise (CCW) or clockwise (CW) edge state current flow. EG on SiC was fabricated into *pn*J devices after the growth that took place at a temperature of 1900 °C. First, chips were diced from 4*H*-SiC(0001) wafers (CREE) [48] and chemically cleaned with a 5:1 diluted solution of hydrofluoric acid and deionized water. Just prior to growth, chips were processed with AZ5214E to utilize polymer-assisted sublimation [52]. Finally, after placing the chips on a polished graphite substrate (SPI Glas 22) [50] silicon-face down, the growth was performed, using an ambient argon environment at 1900 °C with a graphite-lined resistive-element furnace (Materials Research Furnaces Inc.) [48]. Corresponding heating and cooling rates of the furnace were about 1.5 °C/s.

Once grown, EG was assessed with confocal laser scanning, optical, and atomic force microscopy (AFM) [53]. Images acquired from these techniques are provided in Fig. 1 and were verification that homogeneous monolayer EG had successfully covered millimeter-scale areas (see supplementary material for additional AFM images). Next, using Pd and Au as protective layers against organic contamination, photolithographic processes were performed, details of which may be found in other work [31, 47]. Once each Hall bar device was completed, it underwent $Cr(CO)_3$ functionalization to reduce the initially high electron density (due to the buffer layer) to approximately $10^{10}$ cm$^{-2}$ [54-58]. The major final steps included the deposition of S1813 photoresist as a spacer layer for intended *n*-type regions, PMMA/MMA photoresist as an additional spacer, and ZEP520A as a photoactive layer, as described in the literature [47, 59]. Uniformity is also verified with Raman spectra (see supplementary material).

Though AFM images suggest a sloped S1813 spacer layer, the preservation of the *n*-type regions can still be accomplished with thicknesses of the order of 100 nm [49]. Furthermore, the upper bound of the junction width resulting from these photolithographic processes was measured to be approximately 200 nm in another work, rendering them of sufficient sharpness to accommodate edge-state propagation [49]. Ultraviolet (UV) light, with a wavelength of 254 nm, was used to realize *p*-type doping in regions without S1813. The longitudinal resistivity was monitored during periods of UV exposure, and additional information and data on this process can be found in the supplementary material.

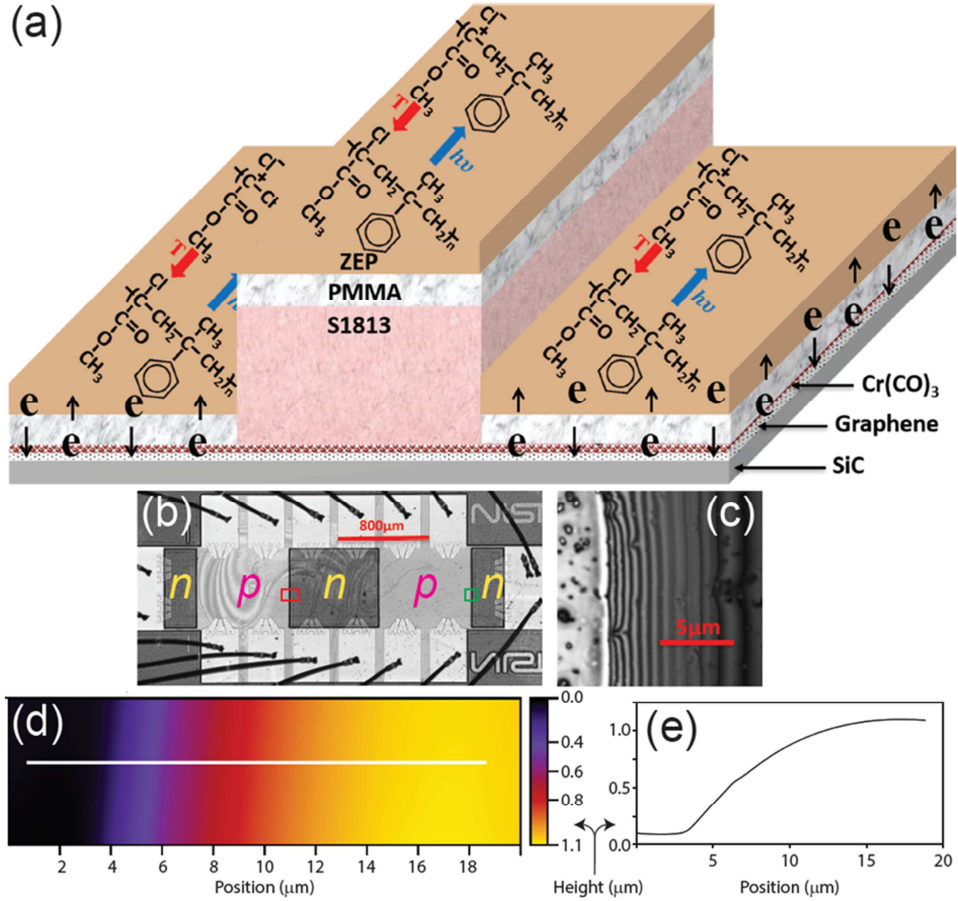

FIG. 1. (a) An illustration of the surface of an EG *pn*J device. The photoresist S1813 was deposited and lithographically processed on specific regions where *n*-type doping was preferred. The molecule in ZEP520A is shown to clarify the electron acceptor as the photoresist is exposed to ultraviolet light. Cr(CO)$_3$ was used to stabilize the electron density. (b) A confocal microscope image of was acquired for the full device after wire bonding, with the darker region indicating the desired *n*-type regions. (c) A magnification is shown of the small green box in (b) for a scale of the order 5 μm. Oxidized residue from the Cr(CO)$_3$ deposition takes the form of visible black specs. (d) and (e) show both the two-dimensional and one-dimensional height profiles, respectively, with the one-dimensional profile represented as a white line in (d) and the two-dimensional profile acquired within the red box in (b).

Completed four-junction devices, like the one shown in Fig. 1 (c), were measured with traditional methods to verify that regions exhibited resistance quantization. The illustration for this type of device is shown in Fig. 2 (a). Electrical contact pads are numbered based on the measurement system used and to provide corresponding measurements in (b) and (c). Traditional longitudinal and Hall measurements were acquired at 1.6 K and ± 9 T, with results shown in Fig. 2 (b) as black and red curves, respectively. With proper UV exposure, regions without the S1813 spacer layer are subject to *p* doping, and after enough exposure time, become set as *p*-type regions.

Resulting *pn*Js were found to be of sufficient narrowness to accommodate dissipationless edge-state propagation [47]. However, to further verify that the entire device was functional, voltage measurements were performed along the length of the device, bearing in mind the formation of the device's so-called hot spots, as shown pictorially by Ref. [40]. In Fig. 2 (c), the plotted resistances further support the idea that millimeter-scale *pn*Js can be successfully fabricated with standard UV lithography.

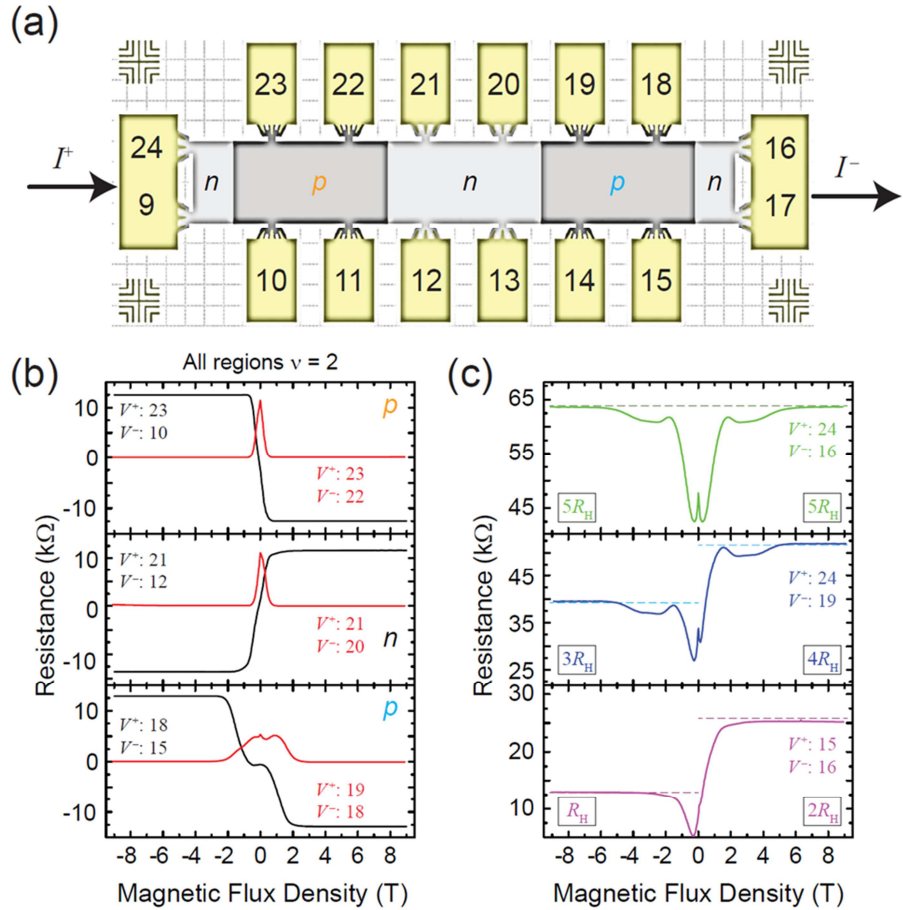

FIG. 2. (a) A four-junction device illustration with numbers corresponding to wired connections on a 32-pin leadless chip carrier, with the electron flow entering from the right-side contact (drain) and with the source on the left side. A current of 1 μA was applied for all measurements. Darker and lighter grey colors indicate *p*-type and *n*-type regions, respectively. The three middle regions are tested to check traditional Hall resistance curves, with orange and cyan *p* labels matching those shown in (b). (b) The longitudinal and Hall resistances were measured from 9 T to -9 T at 1.6 K and are represented by red and black curves, respectively. Pin labels are also provided. (c) Integer multiples of $R_H$ (from 1 to 5) were measured across varying lengths of the device to ensure device functionality. Dotted lines are provided as a visual guide to compare exact quantized values. All regions were on the $\nu = 2$ plateau.

A recent formulation for using multiple terminals on a *pn*J device as the only resistive elements of a circuit has established a mathematical way of predicting the effective quantized resistance of that circuit [49]. Essentially, a single current source can inject current into an arbitrary number of terminals – likewise for the drain port of the current source. The voltage difference of the whole circuit, and by extension the effective quantized resistance $R_{eff} = q_{N-1}R_H$, can then be measured between just after the current source starts and just before the drain of the current source terminates. The coefficient of effective resistance (CER) is labelled as $q$ and represents a device configuration containing $N$ total terminals that are used (either as a source or as a drain).

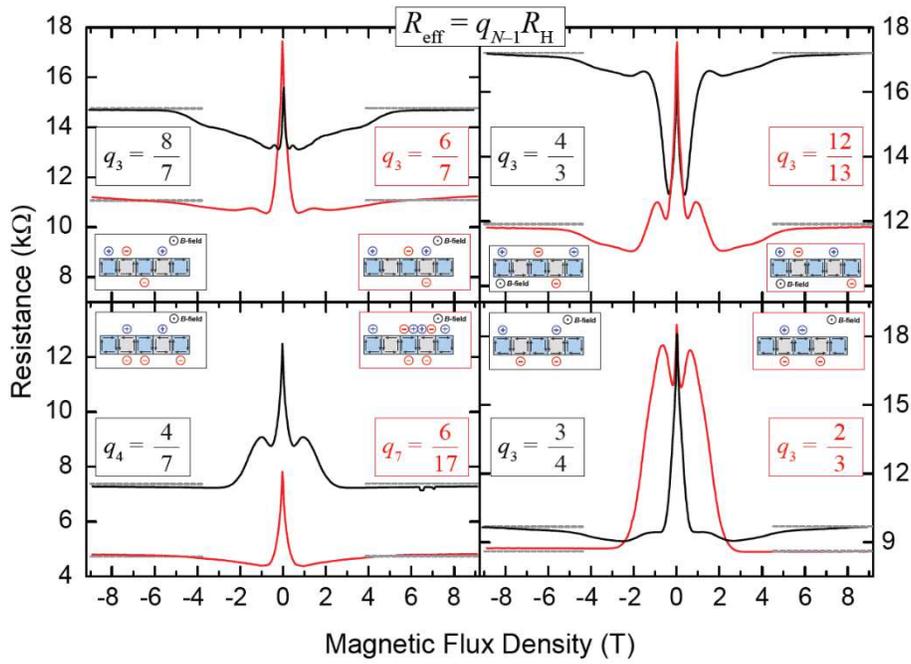

FIG. 3. Multiple-terminal configurations have been measured and their effective circuit resistances are plotted in the four panels above. Three of the four panels contain four-terminal configurations (two sources and two drains) whereas the panel on the lower left corner uses a five- and eight-terminal configuration. The latter panel, when compared with the calculated and simulated value in dotted grey, provides some evidence that the CER formulation is valid for larger numbers of used terminals. All panels contain the calculated and simulated value in dotted grey (valid for sufficiently high magnetic flux density), and for all cases, the calculated and simulated results agree with each other. The insets of each panel have colored perimeters corresponding to the curve of the same color and illustrate the four-junction device and its edge-state current flow abstractly. The blue plus and red minus signs indicate source and drain terminals, respectively.

Eight different configurations were measured, and their effective circuit resistances are plotted in Fig. 3. Furthermore, two methods were used to predict the expected CERs of the circuit – the LTspice simulator and the CER formulation. Both methods agreed exactly and are plotted as grey dotted lines for each of the eight configurations. Eq. (1) below is the crucial formula used to mathematically predict the expected CERs [49]:

$$q_{N-1}(n_{N-1}) = \frac{q_{N-2}(n_{N-1} + 1)}{n_{N-1} + \frac{q_{N-2}}{q_{N-1}^{(0)}}}$$

(1)

The CERs calculated for Fig. 3 include the following: $\left\{ \frac{6}{7}, \frac{8}{7}, \frac{12}{13}, \frac{4}{3}, \frac{6}{17}, \frac{4}{7}, \frac{3}{4}, \frac{2}{3} \right\}$. Details on how to proceed with the calculation are well-documented in Ref. [49], and additional examples for some of the configurations in this manuscript can be found in the supplementary material.

To demonstrate how increasingly complex calculations can yield atypical CERs, devices containing seven *pn*Js were fabricating as shown in Fig. 4 (a). Though even more *pn*Js can be placed along the 2 mm length of the device, their number was limited by the preference of accessing each region with an electrical contact for proof of concept. The illustration in Fig.

4 (a) shows voltage leads of varying color that were used for determining the resistance curves, and by extension the CERs, in Fig. 4 (b).

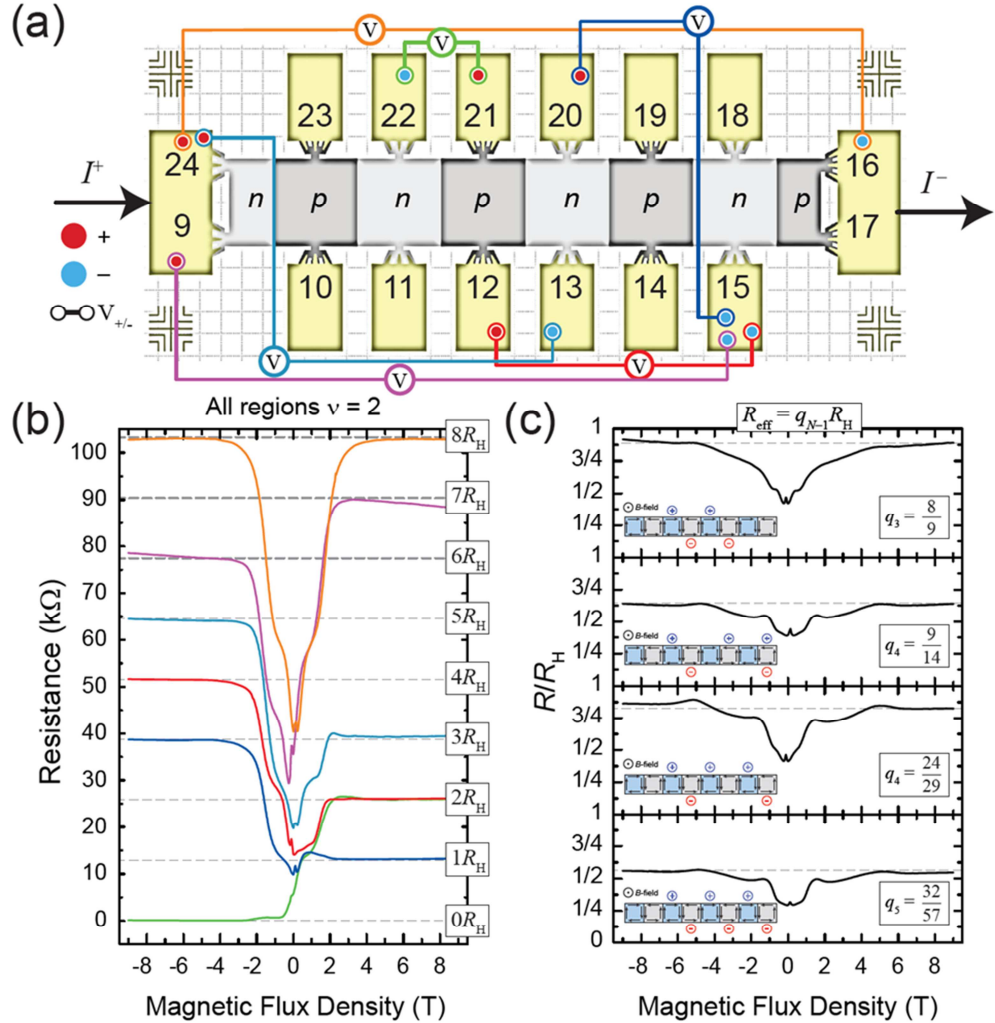

FIG. 4. (a) A seven-junction device illustration with numbers corresponding to the same measurement system is shown using a measurement current of 1 μA. Darker and lighter grey colors indicate $p$-type and $n$-type regions, respectively. Each measurement pair is color-coded for easy comparison to its corresponding data. (b) The resistance is plotted for each of the voltage measurement pairs in (a), with the same color-coding used to match the illustration. The dotted grey lines represent the exact values of the multiples of $R_H$. All regions were on the $\nu = 2$ plateau. (c) Several new configurations were measured and compared with both simulations and the CER formulation, with the latter two agreeing exactly. Thus, both theoretical values are represented by the same grey dotted line in each of the four panels. On the bottom left of each graph panel, the illustrated device is shown with the corresponding locations of sources (blue plus symbol) and drains (red minus symbol) along the device. Voltages were measured from the point before the sources split to the point after the drains rejoin, yielding the CERs of each configuration.

Sufficient quantization was seen for the more traditional cases of measuring the resistance across parts of the device while the source and drain are at the farthest terminals. All integer multiples of $R_H$ between 1 and 8 were accessible in this characterization, warranting further measurements with multiple terminals. In Fig. 4 (c), four configurations were measured using different numbers of total terminals. The top panel, using four terminals as illustrated in the inset, yielded data that was then compared to the predicted CER of $q_3 = \frac{8}{9}$. The two middle panels both used five terminals and were compared with their corresponding predicted values of $q_4 = \frac{9}{14}$ and $q_4 = \frac{24}{29}$. In the bottom panel, the six-terminal configuration was measured and compared with its corresponding prediction of $q_5 = \frac{32}{57}$. For the sake of clarity and as an additional tutorial, this fourth case is calculated in more detail in the supplementary material. Overall, such devices and their CERs can be measured for many configurations of similar or greater complexity. Moreover, desired, user-specific CERs can be reversed engineered into a corresponding configuration.

In conclusion, this work pursued further efforts involving *pn*J devices fabricated from EG on SiC, with junction widths sufficiently narrow to observe usual edge-state propagation. By configuring an experimental setup to include multiple sources and drains, various atypical quantized resistances became accessible and matched predicted values based on LTspice simulations. Additionally, recently reported analytical methods were also used to support the predicted values of the same atypical fractions of $R_H$. The results demonstrate that *pn*Js have the potential to bring scalable resistance values, as well reinforce the validity of the aforementioned CER formulation, which provides a simple algorithm for determining the effective quantized resistances in *pn*J circuits.

## SUPPLEMENTARY MATERIAL

See supplementary material for the details on UV exposure and for additional calculations.

## ACKNOWLEDGMENTS

The work of DKP at NIST was made possible by C-T Liang of National Taiwan University. The work of MM at NIST was made possible by M Ortolano of Politecnico di Torino and L Callegaro of Istituto Nazionale di Ricerca Metrologica, and the authors thank them for this endeavor. The authors would like to express thanks to T Oe and X Wang for their assistance in the NIST internal review process.